\documentclass[aps,nofootinbib, superscriptaddress, notitlepage]{revtex4-1}

\usepackage[mathscr]{eucal}
\usepackage[dvips]{graphicx}
\usepackage{amsfonts,amssymb,amsmath,bbm}
\usepackage[all]{xy}  
\usepackage{tikz}

\newcommand{\be}{\begin{equation}}
\newcommand{\ee}{\end{equation}}
\newcommand{\de}{\partial}
 \usepackage[utf8]{inputenc}
 
\begin{document}

\title{Disformal invariance of continuous media with linear equation of state}
\author{Marco Celoria}
\email{marco.celoria@gssi.infn.it}
\affiliation{Gran Sasso Science Institute (INFN), Viale Francesco Crispi 7,  I-67100 L'Aquila, Italy}

\author{Sabino Matarrese}
\email{sabino.matarrese@pd.infn.it}

\affiliation{Dipartimento di Fisica e Astronomia ``G. Galilei",
Universit\`{a} degli Studi di Padova, via Marzolo 8, I-35131 Padova, Italy}
\affiliation{INFN, Sezione di Padova, via Marzolo 8, I-35131 Padova, Italy}
\affiliation{ INAF-Osservatorio Astronomico di Padova, Vicolo dell' Osservatorio 5, I-35122 Padova, Italy}
\affiliation{Gran Sasso Science Institute (INFN), Viale Francesco Crispi 7, I-67100  L'Aquila, Italy}

\author{Luigi Pilo}
\email{luigi.pilo@aquila.infn.it}
\affiliation{Dipartimento di Fisica, Universit\`{a} di L'Aquila, I-67010 L'Aquila, Italy}
\affiliation{INFN, Laboratori Nazionali del Gran Sasso, I-67010 Assergi, Italy}

\date{\today} 

\begin{abstract}
We show that the effective theory describing single component continuous
media with  a linear and constant equation of state of the form
$p=w\rho$ is invariant under a 1-parameter family of continuous disformal transformations. In the special case  of $w=1/3$
(ultrarelativistic  gas), such a family reduces to 
conformal transformations. As examples,  perfect fluids,  homogeneous and isotropic
solids are discussed.
\end{abstract}

\maketitle

\section{Introduction}

Disformal transformations have been considered in the framework of large-distance modification of gravity \cite{Bekenstein:1992pj}, and have become a very active field of research in modern cosmology and modified gravity, finding applications, for example, in Bekenstein’s TeVeS formalism  \cite{Bekenstein:2004ne}, bimetric theories of gravity \cite{Milgrom:2009gv}, scalar-tensor theories \cite{Clifton:2011jh, Sakstein:2014isa} and disformal inflation \cite{Kaloper:2003yf}. 
Introduced  in the context of Finsler geometry \cite{Bekenstein:1992pj}, these transformations describe the most general relation between two geometries in one and the same gravitational theory, preserving the causal structure and the weak equivalence principle.
\\
More recently, it was discovered that the structure of Horndeski Lagrangian is invariant under a particular class of disformal transformations  \cite{Bettoni:2013diz}. 
\\
As noted in Ref.\cite{Deruelle:2014zza}, the Einstein field equations are invariant under invertible disformal transformations, as well as more general scalar-tensor theories \cite{Arroja:2015wpa}\footnote{This is actually true for a generic non-singular field redefinition}.
This fact has been used  to show that some extensions of Horndeski theory, obtained by performing a  general disformal transformation to the Horndeski Lagrangian, lead to  the same equations of motion of the original theory and are thus free of ghost instabilities \cite{Zumalacarregui:2013pma}. 

On the other hand, if General Relativity (or a more general
scalar-tensor theory) is reformulated in terms of an auxiliary metric,
which is  related with the 
original ``physical" metric by a non-invertible disformal
transformation,  a new  degree
of freedom of gravity is 
switched on \cite{Deruelle:2014zza, Arroja:2015wpa, Arroja:2015yvd}.
 This scalar degree of freedom behaves as an irrotational pressureless
 perfect fluid, i.e. it can mimic a cold dark matter component, or,
 more generally, 
irrotational dust.
  The resulting theory, called mimetic dark matter, was first proposed
  in \cite{Chamseddine:2013kea} and explored in
  \cite{Chamseddine:2014vna} 
(see also \cite{Hammer:2015pcx, Myrzakulov:2015kda, Cognola:2016gjy} and \cite{Ramazanov:2016xhp} for the relation with Ho\v{r}ava-Lifshitz gravity).

Remarkably, disformal transformations have found other applications,
for example they have been used in the context of primordial tensor
modes during  inflation \cite{Creminelli:2014wna}.
\\
Furthermore,  in a general fixed background  even a free massless
scalar field  is invariant under a particular class of local disformal
transformations~\cite{Falciano:2011rf} and disformal invariance of
Maxwell's field equations was proved in \cite{Goulart:2013laa} while 
invariance of the Dirac equation was studied in~\cite{Bittencourt:2015ypa}.

The goal of this paper is to analyze the general consequences of
invariance under a disformal transformation for a  rather large class of systems, namely
continuous media, that are relevant for  cosmology, modified gravity or analog gravity.

Our analysis relies on the  observation that  the physics of  fluids
or solids can be derived by an unique Lagrangian \cite{Taub:1954zz, Schutz:1970my, Schutz:1977df}.
This remarkable result has been investigated using the language of the
effective field theory \cite{Dubovsky:2005xd,Dubovsky:2011sj,Nicolis:2011cs,Ballesteros:2012kv,Delacretaz:2014jka,Gripaios:2014yha}
using the pull-back formalism, which had already been used to describe
the dynamics of continuous media \cite{Carter:1987qr,Comer:1993zfa,Comer:1994tw,Andersson:2006nr}.
Basically, the  symmetries of  the theory allow to extract all the dynamical  information and the thermodynamics
concerning the medium from an action principle.

The paper is organized as follows. 
In Section \ref{Section:GeneralDisformal} we show that any  Lagrangian  invariant under this class of disformal transformations describes a medium with a linear equation of state. 
Then, as an example, in Section  \ref{Section:PFDisformal} we show
that the action for a  perfect fluid with a linear equation of state
is disformal invariant, the special case of an 
irrotational perfect fluid is discussed in Section \ref{Section:IPFDisformal}.
In Section \ref{Section:SolidDisformal}  we generalize this result to homogeneous and isotropic solids. 
Finally, in Section \ref{Section:Discussion}, we summarize and discuss our results.

\textit{Notation}. Our signature is $(-+++)$; greek indices run from $0$ to $3$. Units are such that $\hbar=c=k_B=1$.

\section{Weyl and Disformal transformations}\label{Section:GeneralDisformal}
Invariance under Weyl (or conformal) transformations  is characterized by a vanishing
trace of the energy-momentum tensor (EMT), disformal ones constitute
a generalization.

Let us start by considering the rather  general action 
\begin{equation}\label{eq:generic_action}
S
=\int d^4 x\sqrt{-g} \,\mathcal{L}(g_{\mu\nu},\phi^I, \partial
\phi^I,\partial_1\dots \partial_n \phi^I ) \, ;
\end{equation}
 where $I=1,\dots, N $ and $\phi^I$ are generic matter fields. 
A Weyl transformation is  defined by
\begin{equation}\label{eq:Weyl_transformation}
g_{\mu\nu}\rightarrow \bar{g}_{\mu\nu}=e^{\sigma(x)}g_{\mu\nu} \, ;
\end{equation}
or  considering infinitesimal transformations connected with
identity, corresponding  to $\sigma_0=0$, we get 
\begin{equation}
\delta g_{\mu\nu} = \bar{g}_{\mu\nu}-g_{\mu\nu}=g_{\mu\nu}
\delta\sigma +\mathcal{O}(\delta\sigma ^2) \, .
\end{equation}
If the action  (\ref{eq:generic_action}) is invariant under (\ref{eq:Weyl_transformation}), then 
\begin{equation}
\delta S =\frac{1}{2}\int  d^4x \sqrt{-g} T^{\mu\nu}g_{\mu\nu}\delta\sigma
 = 0   \, ;
\end{equation}
 we then conclude that  the trace of the EMT vanishes
\begin{equation}
T^\mu_{\phantom{\mu}\mu}=0 \, .
\end{equation}
The extension to one parameter group of disformal  transformations
connected to the identity crucially depends on the
the matter content. In particular, a 4-vector emerging from the
matter's action is needed. The simplest option is to 
suppose that the energy-momentum tensor describes  a single-component
medium  of the following form~\cite{Andersson:2006nr} 
\begin{equation}
T_{\mu\nu} =  \rho \, u_\mu \, u_\nu + p \, h_{\mu\nu}  
+ \pi_{\mu\nu}  \; ;
\label{EMT}
\end{equation}
where $u_\mu$ is the timelike eigenvector of $T_{\mu\nu}$ and
$h_{\mu\nu} = g_{\mu\nu} + u_{\mu} \, u_{\nu}$ is the  projector
orthogonal to $u_\mu$. The energy density $\rho$,  the isotropic pressure $p$ and the anisotropic stress $\pi_{\mu\nu}$ are respectively defined as the
following projections of the energy-momentum tensor $T_{\mu \nu}$
\begin{align}
\rho = T_{\mu\nu}u^\mu u^\nu, 
\qquad  p= \frac{1}{3} T_{\mu\nu}h^{\mu\nu} , 
\qquad \pi_{\mu\nu} = h^{\alpha}_{\phantom{\alpha}\mu}h^{\sigma}_{\phantom{\sigma}\nu}T_{\alpha\sigma} - \frac{1}{3}h_{\mu\nu} h_{\alpha\sigma}T^{\alpha\sigma} . 
\end{align}
We have also supposed that the EMT is in the so-called Landau-Lifshitz frame~
\cite{Kovtun:2012rj, landau1959fluid}.   
By construction the anisotropic tensor is
traceless, $\pi^{\mu}_{\phantom{\mu}\mu}=0$. The key property needed to
define disformal transformation is the existence of of a single
timelike eigevenvector $u_\mu$.

Consider now the one-parameter group of disformal transformations defined by
\begin{equation}
\begin{aligned}\label{eq:generic_disformal_transformation}
g_{\mu\nu}\rightarrow \bar{g}_{\mu\nu}
&=e^{\sigma}g_{\mu\nu}+ \left[e^{\sigma}-e^{3w\sigma}\right] u_{\mu}u_{\nu}
\\
&=
e^{\sigma}h_{\mu\nu}-e^{3w\sigma} u_{\mu}u_{\nu}
\\
 &\equiv
 Ah_{\mu\nu} -Cu_{\mu}u_{\nu}
\end{aligned} 
\end{equation}
so that for $\sigma=0$ we obtain the identity
transformation, and the factor $3$ is included for later convenience. Expanding around the identity, we get the variation of the metric 
\begin{equation}
\delta g_{\mu\nu}=\bar{g}_{\mu\nu}
-g_{\mu\nu}=\left[g_{\mu\nu}+\left(1
    -3w\right)u_\mu
  u_\nu  \right]\delta\sigma \, . 
\end{equation}
Suppose that the action  (\ref{eq:generic_action}),  when the EMT has
the form (\ref{EMT}),  is invariant under the transformation (\ref{eq:generic_disformal_transformation})
\begin{equation}
\delta S=
\frac{1}{2}\int  d^4x \sqrt{-g} T^{\mu\nu}\delta g_{\mu\nu}
=
\frac{1}{2}\int  d^4x \left[T^{\mu}_{\phantom{\mu}\mu}+\left(1
    -3w
  \right)\rho \right]\delta\sigma =0\; .
\end{equation}
Therefore,  we have
\begin{equation}
3 \, p -\rho
=\left(3w
  -1\right)\rho \, ;
\end{equation}
and we conclude that one-parameter disformal transformation
(\ref{eq:generic_disformal_transformation}) can be defined for
one-component fluids and invariance  implies
\begin{itemize}
\item  a linear equation of state
$p(\rho)=w \rho$; 
\item a constant speed of sound $c_s^2=w$.
\end{itemize}
Note that for $w=\frac{1}{3}$, the  disformal transformation (\ref{eq:generic_disformal_transformation}) reduces to
a   Weyl transformation.

\section{Perfect fluids and Disformal Transformations}\label{Section:PFDisformal}
The results of the previous section are very general, here we focus on the case of perfect fluids.
The most general class of perfect fluids can be effectively described
by using four scalar fields $\Phi^A$ representing Eulerian
coordinates on the fluid's world-volume, or, equivalently,  as
Stueckelberg fields  associated  to broken space-time translations. 
At leading order (LO) in a derivative expansion the action can be written
as ~\cite{Ballesteros:2016gwc, Ballesteros:2016kdx, Dubovsky:2011sj}
\begin{equation}
S=\int d^4x\sqrt{-g} \, U(b,Y)
\end{equation}
 where $U$ is a generic function and 
 \begin{equation}
 B^{mn}=
 g^{\mu\nu}\partial_\mu\Phi^m(x,\tau)\partial_\nu\Phi^n(x,\tau),
 \qquad m,n=1,2,3 , \, \qquad b=\sqrt{\text{det}(\pmb B)} \, , \qquad 
 Y=u^\mu\partial_\mu\Phi^0(x,\tau) ;
 \end{equation}
the fluid's 4-velocity satisfies
 \begin{equation}
 \label{eq:fluidvelocity}
 u^\mu \partial_\mu \Phi^a=0\,,\qquad u^\mu u_\mu=-1\,.
 \end{equation}
The perfect fluid Lagrangian is invariant under
$\Phi^0\rightarrow \Phi^0+f(\Phi^a)$
and internal volume-preserving diffeomorphisms $\Phi^a\longrightarrow
f^a(\Phi)$ with $\det\left(\frac{\partial f^a}{\partial \Phi^b}\right)=1$.
The  energy-momentum tensor of the system is
 of course the one of perfect fluid with \cite{ Dubovsky:2011sj, Ballesteros:2016kdx}
\begin{equation}
\rho=Y\, U_{Y} -U \, , \qquad p=U-b\,U_{b}\,. 
\end{equation}
Notice that in general the fluid equation of state is not barotropic,
unless a special form for $U$ is chosen. 
For instance, a  barotropic equation of state for $U(Y,b)$ with $p/\rho=
w =$const. can be obtained by choosing \cite{Ballesteros:2016gwc} $U\propto b^{1+w} \mathcal{U} ( b^{-w} Y )$
where $\mathcal{U}$ is an arbitrary function or alternatively
 $U\propto (b^{1+w} + Y ^{1+1/w} )$.
Dust (pressureless matter) $p=0$  is obtained when  $U \propto
\mathcal{U}(Y)b$, while radiation  $p=\frac{1}{3}\rho$ when  $U\propto b^{4/3} \mathcal{U} ( b^{-1/3} Y )$. 
Finally, an equation of state $w = -1$ can be obtained by choosing $U \propto \mathcal{U} (b Y )$.

From the general discussion in section \ref{Section:GeneralDisformal},
  the action describing a perfect fluid with a linear equation of
  state  $p=w\rho$ is invariant under a disformal
  transformation. Indeed, under
  (\ref{eq:generic_disformal_transformation}) we have
\begin{equation}
B^{ab}\rightarrow \bar{B}^{ab}=B^{ab}\,A^{-1} \, , \qquad b
\rightarrow \bar{b}= b\,A^{-3/2} \, , \qquad Y  \rightarrow \bar{Y}=
Y\, C^{-1/2} \, .
\label{Btrans}
\end{equation}
The  determinant of the metric transforms as
 \begin{equation}
 \sqrt{-g} \, \rightarrow \, \sqrt{-\bar{g}} \,= \, \sqrt{-g} \, A^{3/2}\sqrt{C}
 \end{equation}
 (see Appendix C of   \cite{Bekenstein:2004ne}  for a derivation of the above expression), so we conclude that the action transforms as
\begin{equation}
S=\int d^4x\sqrt{-g}\,U(b,Y)\rightarrow \int d^4x \sqrt{-g}\,A^{3/2}\sqrt{C} \,U(b\,A^{-3/2}, Y\,C^{-1/2})\,.
\end{equation}
By considering an infinitesimal transformation,  $|\delta\sigma|\ll 1$, it is easy to show that the action is invariant if
\begin{align}
U
  -b\,U_{,b}=w\left(Y\,U_{Y}-
  U\right) \, .
\end{align}
As a result 
\begin{equation}
p=w\rho
\end{equation}
i.e. a barotropic perfect fluid characterized by a linear equation of
state with $w$ constant.
The case  $w=-1$, i.e. a cosmological constant, and $w=0$, i.e. non relativistic matter (dust), are subcases of a perfect fluid with $p= -\rho$ and $p=0$, respectively. 

\section{Irrotational Perfect Fluids}\label{Section:IPFDisformal}
At  leading order in a derivative expansion the most general
action describing a fluid can be written as~\cite{Dubovsky:2011sj,Nicolis:2011cs,Ballesteros:2016gwc,Ballesteros:2016kdx}
 \begin{equation}
S=\int d^4x\sqrt{-g}\,U(b,Y,X) \, ;
\end{equation}
with
 \begin{equation}
 X=g^{\mu\nu}\partial_\mu \Phi^0\partial_\nu\Phi^0 \, .
 \end{equation}
Actually, the EMT derived from the above action describes a
two-component fluid system or a superfluid. An interesting subcase is when
only $X$ is present, namely $U(X)$ and it was used~\cite{Matarrese:1984zw} as one of the first examples of the effective
description of fluid dynamics in terms of scalar fields.    
Here, we present the effective description of irrotational
perfect fluids and irrotational dust, and we show that the invariance
under disformal transformation implies a linear equation of state. 
The EMT reads
 \begin{equation}
 T_{\mu\nu} =
 2\, U' \, X \, v_{\mu} \, v_\nu + U \, g_{\mu\nu} \, , \qquad v^{\mu}=\frac{\partial^\mu\Phi^0}{\sqrt{-X}}\, ; 
 \end{equation}
which has  a perfect fluid form with
 \begin{equation}
 \rho = 2\, U' \, X - U, \qquad p=U\,.
 \end{equation}
In  addition being the 4-velocity $v_\mu$ proportional to an exact
1-form, the fluid is irrotational. 

By performing an infinitesimal disformal transformation, it is possible to show that the action is invariant if \footnote{More generally, we can have $X^\alpha f(\Phi^0)$, with $f$ an arbitrary function.}
 \begin{equation}
 U(X)\sim X^{\frac{1}{2}\left(1 +\frac{1}{w}\right)}
 \end{equation}
corresponding to an irrotational perfect fluid characterized by a linear equation of state, $p=w\rho$. 
\\
Note that the effective Lagrangian is singular for $w=0$. In this case, we consider the mimetic action \cite{Brown:1994py, Chamseddine:2013kea}
\begin{equation}
S=\int d^4x\sqrt{-g}\,\rho(x)\left(X+1\right)
\end{equation}
which describes irrotational fluid-like dust with the energy density
given by the Lagrange multiplier $\rho$ and velocity potential
$\Phi^0$. 
 The energy-momentum tensor is
 \begin{equation}
 \begin{aligned}
 T_{\mu\nu}^{on-shell}&=\rho\,\partial_\mu\Phi^0\partial_\nu\Phi^0 \, ;
 \end{aligned}
 \end{equation}
 where we identify $\rho$ as the on-shell energy density, e.g. when
 the constraint $X=-1$ is enforced.
 
 Moreover, the theory is invariant under the shift symmetry $\Phi^0\rightarrow \Phi^0+\lambda$, with $\lambda=const.$. The associated conserved Noether current  is 
 \begin{equation}
 J_\mu=\rho\, \partial_\mu\Phi^0 \, .
 \end{equation}
 The mimetic action is invariant under  (\ref{eq:generic_disformal_transformation}) with $w=0$, i.e. a conformal transformation of the spatial metric
 \begin{equation}
 \begin{cases}\label{eq:disfw0bis}
 h_{\mu\nu}&\rightarrow \bar{h}_{\mu\nu}=A h_{\mu\nu}\\
 \rho&\rightarrow \bar{\rho}=A^{-\frac{3}{2}}\rho
   \end{cases} \, .
 \end{equation}
where the rescaling of the  Lagrange multiplier $\rho$ is required by the conservation of the Noether current under (\ref{eq:generic_disformal_transformation}), $\nabla_\mu J^\mu =0 \rightarrow \bar{\nabla}_\mu \bar{J}^\mu=0$.

\section{Solids and Disformal transformations}\label{Section:SolidDisformal}
Finally,  let us  consider the case of isotropic and homogeneous solids, characterized by the presence of a non vanishing anisotropic tensor, and described at LO by the effective action  \cite{Endlich:2012pz, Ballesteros:2016gwc}  
\begin{equation}
S=\int d^4x \sqrt{-g}\, U(\tau_1,\tau_2,\tau_3) \, ;
\end{equation}
where 
\begin{align}
\tau_n =\text{Tr}({\pmb B}^n)\,,\quad n=1,2,3\, \; , \qquad
  \left[\pmb B \right]^{ab} = B^{ab} \,.
\end{align}
The EMT has the following form 
\be
T_{\mu \nu} = U \,  g_{\mu \nu}  -2\left(U_{\tau_1} \, \delta^{ab} + 2 
\, U_{\tau_2} \, B^{ab} + 3 \, U_{\tau_3} \,  B^{ac} \, B^{cb}\right)\de_\mu \Phi^a \de_\nu
\Phi^b\,,
\label{tela}
\ee
and represents 
the relativistic generalization of the energy-momentum tensor of an elastic
material.  The stress state is given by $B^{ab}$ and the isotropic solid or
jelly is not stressed when $B^{ab} = \delta^{ab}$. In particular
\begin{align}
\rho = -U\,,\quad p=U-2\sum_{n=1}^3 n\, U_{\tau_n}\tau_n\,.
\end{align}
In the previous sections, we have seen that  a perfect fluid, characterized by a linear equation of state $p=w\rho$, is invariant under the disformal transformation (\ref{eq:generic_disformal_transformation}).
Conversely, a perfect fluid, invariant under this disformal transformation, is necessarily  described by a linear equation of state.
Here we  generalize this invariance property to solids with linear
equation of state. From (\ref{Btrans}) we have that $\tau_n \to A^{-n}
\, \tau_n$, $n=1,2,3$.
Thus the solid action transforms as 
\begin{equation}
S=\int d^4x\sqrt{-g}\, U(\tau_1, \tau_2,\tau_3)\rightarrow \bar{S}=\int
d^4x \sqrt{-g} \, A^{3/2}  \sqrt{C}\, U(A^{-1} \tau_1, \, A^{-2} \tau_2,
A^{-3} \tau_3)  \,.
\end{equation}
Under an infinitesimal transformation  $|\delta\sigma|\ll 1$,  the action is invariant if
\begin{align}
3 (1+ w) \, U - \sum_{n=1}^3 n \, U_{\tau_n} =0  \, .
\label{eqin}
\end{align}
 The general solution of (\ref{eqin}) can
be written as
\begin{equation}
U= \tau_1^{3(1+w)/2} \, {\cal F}\left(\frac{\tau_2}{\tau_1^2}\, ,
\frac{\tau_3}{\tau_1^3}\right) \, , 
\end{equation}
which leads to a  barotropic equation of state of the form:  $p= (2+ 3 \, w) \rho$. 
Thus, the action for a solid, characterized by a linear equation of state, is invariant under (\ref{eq:generic_disformal_transformation}).

\section{Discussion}\label{Section:Discussion}
In this work we have studied the invariance properties under a
1-parameter family class of disformal transformations (\ref{eq:generic_disformal_transformation}), that can be considered as a deformation of Weyl transformations.
\\
Each family is characterized by a constant $w$ that has a manifest physical interpretation.
Indeed, we have shown that every Lagrangian invariant under these
disformal transformations is associated with an energy-momentum tensor
characterized by a linear and  equation of 
state of the form  $p=w\rho$. 
\\
When  $w=1/3$ the family of disformal transformations reduces to
conformal ones and we recover the well-known conformal invariance of a 
Lagrangian describing ultra-relativistic matter.
As explicit examples of our general analysis we have considered
perfect fluids and solids, generalizing the analysis for a  free
scalar field given in \cite{Falciano:2011rf}.

\section{Acknowledgments}
We are indebted to Sabir Ramazanov for many useful comments and discussions. We also thank the Galileo Galilei Institute for Theoretical Physics (GGI) in Florence for hospitality and the Istituto Nazionale di Fisica Nucleare (INFN) for support during the completion of this work.

\bibliographystyle{hunsrt}  
  
\bibliography{fluidbiblio}

\begin{thebibliography}{10}

\bibitem{Bekenstein:1992pj}
Jacob~D. Bekenstein.
\newblock {The Relation between physical and gravitational geometry}.
\newblock {\em Phys. Rev.}, D48:3641--3647, 1993, gr-qc/9211017.

\bibitem{Bekenstein:2004ne}
Jacob~D. Bekenstein.
\newblock {Relativistic gravitation theory for the MOND paradigm}.
\newblock {\em Phys. Rev.}, D70:083509, 2004, astro-ph/0403694.
\newblock [Erratum: Phys. Rev.D71,069901(2005)].

\bibitem{Milgrom:2009gv}
Mordehai Milgrom.
\newblock {Bimetric MOND gravity}.
\newblock {\em Phys. Rev.}, D80:123536, 2009, 0912.0790.

\bibitem{Clifton:2011jh}
Timothy Clifton, Pedro~G. Ferreira, Antonio Padilla, and Constantinos Skordis.
\newblock {Modified Gravity and Cosmology}.
\newblock {\em Phys. Rept.}, 513:1--189, 2012, 1106.2476.

\bibitem{Sakstein:2014isa}
Jeremy Sakstein.
\newblock {Disformal Theories of Gravity: From the Solar System to Cosmology}.
\newblock {\em JCAP}, 1412:012, 2014, 1409.1734.

\bibitem{Kaloper:2003yf}
Nemanja Kaloper.
\newblock {Disformal inflation}.
\newblock {\em Phys. Lett.}, B583:1--13, 2004, hep-ph/0312002.

\bibitem{Bettoni:2013diz}
Dario Bettoni and Stefano Liberati.
\newblock {Disformal invariance of second order scalar-tensor theories: Framing
  the Horndeski action}.
\newblock {\em Phys. Rev.}, D88:084020, 2013, 1306.6724.

\bibitem{Deruelle:2014zza}
Nathalie Deruelle and Josephine Rua.
\newblock {Disformal Transformations, Veiled General Relativity and Mimetic
  Gravity}.
\newblock {\em JCAP}, 1409:002, 2014, 1407.0825.

\bibitem{Arroja:2015wpa}
Frederico Arroja, Nicola Bartolo, Purnendu Karmakar, and Sabino Matarrese.
\newblock {The two faces of mimetic Horndeski gravity: disformal
  transformations and Lagrange multiplier}.
\newblock {\em JCAP}, 1509:051, 2015, 1506.08575.

\bibitem{Zumalacarregui:2013pma}
Miguel Zumalacárregui and Juan García-Bellido.
\newblock {Transforming gravity: from derivative couplings to matter to
  second-order scalar-tensor theories beyond the Horndeski Lagrangian}.
\newblock {\em Phys. Rev.}, D89:064046, 2014, 1308.4685.

\bibitem{Arroja:2015yvd}
Frederico Arroja, Nicola Bartolo, Purnendu Karmakar, and Sabino Matarrese.
\newblock {Cosmological perturbations in mimetic Horndeski gravity}.
\newblock {\em JCAP}, 1604(04):042, 2016, 1512.09374.

\bibitem{Chamseddine:2013kea}
Ali~H. Chamseddine and Viatcheslav Mukhanov.
\newblock {Mimetic Dark Matter}.
\newblock {\em JHEP}, 11:135, 2013, 1308.5410.

\bibitem{Chamseddine:2014vna}
Ali~H. Chamseddine, Viatcheslav Mukhanov, and Alexander Vikman.
\newblock {Cosmology with Mimetic Matter}.
\newblock {\em JCAP}, 1406:017, 2014, 1403.3961.

\bibitem{Hammer:2015pcx}
Katrin Hammer and Alexander Vikman.
\newblock {Many Faces of Mimetic Gravity}.
\newblock 2015, 1512.09118.

\bibitem{Myrzakulov:2015kda}
Ratbay Myrzakulov, Lorenzo Sebastiani, Sunny Vagnozzi, and Sergio Zerbini.
\newblock {Static spherically symmetric solutions in mimetic gravity: rotation
  curves and wormholes}.
\newblock {\em Class. Quant. Grav.}, 33(12):125005, 2016, 1510.02284.

\bibitem{Cognola:2016gjy}
Guido Cognola, Ratbay Myrzakulov, Lorenzo Sebastiani, Sunny Vagnozzi, and
  Sergio Zerbini.
\newblock {Covariant renormalizable gravity model as a mimetic Horndeski model:
  cosmological solutions and perturbations}.
\newblock 2016, 1601.00102.

\bibitem{Ramazanov:2016xhp}
S.~Ramazanov, F.~Arroja, M.~Celoria, S.~Matarrese, and L.~Pilo.
\newblock {Living with ghosts in Hořava-Lifshitz gravity}.
\newblock {\em JHEP}, 06:020, 2016, 1601.05405.

\bibitem{Creminelli:2014wna}
Paolo Creminelli, Jérôme Gleyzes, Jorge Noreña, and Filippo Vernizzi.
\newblock {Resilience of the standard predictions for primordial tensor modes}.
\newblock {\em Phys. Rev. Lett.}, 113(23):231301, 2014, 1407.8439.

\bibitem{Falciano:2011rf}
F.~T. Falciano and E.~Goulart.
\newblock {A new symmetry of the relativistic wave equation}.
\newblock {\em Class. Quant. Grav.}, 29:085011, 2012, 1112.1341.

\bibitem{Goulart:2013laa}
E.~Goulart and F.~T. Falciano.
\newblock {Disformal invariance of Maxwell's field equations}.
\newblock {\em Class. Quant. Grav.}, 30:155020, 2013, 1303.4350.

\bibitem{Bittencourt:2015ypa}
Eduardo Bittencourt, Iarley~P. Lobo, and Gabriel~G. Carvalho.
\newblock {On the disformal invariance of the Dirac equation}.
\newblock {\em Class. Quant. Grav.}, 32:185016, 2015, 1505.03415.

\bibitem{Taub:1954zz}
A.~H. Taub.
\newblock {General Relativistic Variational Principle for Perfect Fluids}.
\newblock {\em Phys. Rev.}, 94:1468--1470, 1954.

\bibitem{Schutz:1970my}
Bernard~F. Schutz.
\newblock {Perfect Fluids in General Relativity: Velocity Potentials and a
  Variational Principle}.
\newblock {\em Phys. Rev.}, D2:2762--2773, 1970.

\bibitem{Schutz:1977df}
Bernard~F. Schutz and Rafael Sorkin.
\newblock {Variational aspects of relativistic field theories, with application
  to perfect fluids}.
\newblock {\em Annals Phys.}, 107:1--43, 1977.

\bibitem{Dubovsky:2005xd}
S.~Dubovsky, T.~Gregoire, A.~Nicolis, and R.~Rattazzi.
\newblock {Null energy condition and superluminal propagation}.
\newblock {\em JHEP}, 03:025, 2006, hep-th/0512260.

\bibitem{Dubovsky:2011sj}
Sergei Dubovsky, Lam Hui, Alberto Nicolis, and Dam~Thanh Son.
\newblock {Effective field theory for hydrodynamics: thermodynamics, and the
  derivative expansion}.
\newblock {\em Phys. Rev.}, D85:085029, 2012, 1107.0731.

\bibitem{Nicolis:2011cs}
Alberto Nicolis.
\newblock {Low-energy effective field theory for finite-temperature
  relativistic superfluids}.
\newblock 2011, 1108.2513.

\bibitem{Ballesteros:2012kv}
Guillermo Ballesteros and Brando Bellazzini.
\newblock {Effective perfect fluids in cosmology}.
\newblock {\em JCAP}, 1304:001, 2013, 1210.1561.

\bibitem{Delacretaz:2014jka}
Luca~V. Delacr\'etaz, Alberto Nicolis, Riccardo Penco, and Rachel~A. Rosen.
\newblock {Wess-Zumino Terms for Relativistic Fluids, Superfluids, Solids, and
  Supersolids}.
\newblock {\em Phys. Rev. Lett.}, 114(9):091601, 2015, 1403.6509.

\bibitem{Gripaios:2014yha}
Ben Gripaios and Dave Sutherland.
\newblock {Quantum Field Theory of Fluids}.
\newblock {\em Phys. Rev. Lett.}, 114(7):071601, 2015, 1406.4422.

\bibitem{Carter:1987qr}
Brandon Carter.
\newblock {Covariant Theory of Conductivity in Ideal Fluid or Solid Media}.
\newblock {\em Lect. Notes Math.}, 1385:1--64, 1989.

\bibitem{Comer:1993zfa}
G.~L. Comer and D.~Langlois.
\newblock {Hamiltonian formulation for multi-constituent relativistic perfect
  fluids}.
\newblock {\em Class. Quant. Grav.}, 10:2317--2327, 1993.

\bibitem{Comer:1994tw}
G.~L. Comer and D.~Langlois.
\newblock {Hamiltonian formulation for relativistic superfluids}.
\newblock {\em Class. Quant. Grav.}, 11:709--721, 1994.

\bibitem{Andersson:2006nr}
N.~Andersson and G.~L. Comer.
\newblock {Relativistic fluid dynamics: Physics for many different scales}.
\newblock {\em Living Rev. Rel.}, 10:1, 2007, gr-qc/0605010.

\bibitem{Kovtun:2012rj}
Pavel Kovtun.
\newblock {Lectures on hydrodynamic fluctuations in relativistic theories}.
\newblock {\em J. Phys.}, A45:473001, 2012, 1205.5040.

\bibitem{landau1959fluid}
L.D. Landau and E.M. Lifshits.
\newblock {\em Fluid Mechanics, by L.D. Landau and E.M. Lifshitz}.
\newblock Pergamon Press, 1959.

\bibitem{Ballesteros:2016gwc}
Guillermo Ballesteros, Denis Comelli, and Luigi Pilo.
\newblock {Massive and modified gravity as self-gravitating media}.
\newblock 2016, arXiv:1603.02956.

\bibitem{Ballesteros:2016kdx}
Guillermo Ballesteros, Denis Comelli, and Luigi Pilo.
\newblock {Thermodynamics of perfect fluids from scalar field theory}.
\newblock {\em Phys. Rev.}, D94(2):025034, 2016, 1605.05304.

\bibitem{Matarrese:1984zw}
S.~Matarrese.
\newblock {On the Classical and Quantum Irrotational Motions of a Relativistic
  Perfect Fluid. 1. Classical Theory}.
\newblock {\em Proc. Roy. Soc. Lond.}, A401:53--66, 1985.

\bibitem{Brown:1994py}
J.~David Brown and Karel~V. Kuchar.
\newblock {Dust as a standard of space and time in canonical quantum gravity}.
\newblock {\em Phys. Rev.}, D51:5600--5629, 1995, gr-qc/9409001.

\bibitem{Endlich:2012pz}
Solomon Endlich, Alberto Nicolis, and Junpu Wang.
\newblock {Solid Inflation}.
\newblock {\em JCAP}, 1310:011, 2013, 1210.0569.

\end{thebibliography}

\end{document}